\documentclass[fleqn]{article}
\usepackage{amsmath,amssymb}
\usepackage{epsfig,graphicx}

\renewcommand{\theequation}{\arabic{section}.\arabic{equation}}
\renewcommand{\thesection}{\arabic{section}.}

\catcode`@=11 \@addtoreset{equation}{section} \catcode`@=12

\begin{document}
\title{\vskip-1.7cm \bf  On the functional determinant of a special operator
with a zero mode in cosmology}
\date{}
\author{A.O.Barvinsky$^{1,\,2}$ and A.Yu.Kamenshchik$^{3,\,4}$}
\maketitle
\hspace{-8mm} {\,\,$^{1}$\em Theory Department, Lebedev
Physics Institute, Leninsky Prospect 53, Moscow 119991, Russia\\
$^{2}$Department of Physics, Ludwig Maximilians University,
Theresienstrasse 37, Munich, Germany\\
$^{3}$Dipartimento di Fisica and INFN, via Irnerio 46, 40126
Bologna, Italy\\
$^{4}$L.D.Landau Institute for Theoretical Physcis, Kosygin str. 2,
119334 Moscow, Russia}
\begin{abstract}
The functional determinant of a special second order
quantum-mechanical operator is calculated with its zero mode gauged
out by the method of the Faddeev-Popov gauge fixing procedure. This
operator subject to periodic boundary conditions arises in
applications of the early Universe theory and, in particular,
determines the one-loop statistical sum in quantum cosmology
generated by a conformal field theory (CFT). The calculation is done
for a special case of a periodic zero mode of this operator having
two roots (nodes) within the period range, which corresponds to the
class of cosmological instantons in the CFT driven cosmology with
one oscillation of the cosmological scale factor of its Euclidean
Friedmann-Robertson-Walker metric.
\end{abstract}

\section{Introduction}

The theory of long-wavelength perturbations in early Universe
cosmology, including the formation of observable CMB spectra
\cite{MFB,Mukhanov}, essentially relies on the differential equation
with the operator of the form
    \begin{eqnarray}
    \mbox{\boldmath${F}$}=-\frac1{g}\frac{d}{d\tau}
    g^2\frac{d}{d\tau}\frac1{g}
    =-\frac{d^2}{d\tau^2}+\frac{\ddot g}g, \label{operator}.
    \end{eqnarray}
where $g=g(\tau)$ is a rather generic function of the cosmic time
depending on the behavior of the cosmological scale factor $a$ and
its time derivative, $g\varpropto\dot a$. In particular, for
superhorizon cosmological perturbations of small momenta $k^2\ll
\ddot g/g$ their evolution operator only slightly differs from
(\ref{operator}) by adding $k^2$ to its potential term, whereas in
the minisuperspace sector of cosmology, corresponding to spatially
constant variables, the operator has exactly the above form.

Up to an overall sign, this operator is the same both in the
Lorentzian and Euclidean signature spacetimes with the time
variables related by  the Wick rotation $\tau=it$. In the Euclidean
case it plays a very important role in the calculation of the
statistical sum for the microcanonical ensemble in cosmology -- a
concept of initial conditions, which is very promising from the
viewpoint of the cosmological constant, inflation and dark energy
problems \cite{slih,why,bigboost,tunnel,PIQC}. However, in this
statistical theory context, when $\tau$ plays the role of the
Euclidean time, the properties of this operator essentially differ
from the Lorentzian dynamics. In the latter case the function $g$ is
a monotonic function of time because of the monotonically growing
cosmological scale factor, whereas in the Euclidean case $g(\tau)$
is periodic just as the scale factor $a(\tau)$ itself and, moreover,
has zeroes at turning points of the Euclidean evolution with $\dot
a=0$, because $g(\tau)\varpropto\dot a(\tau)$.

This does not lead to a singular behavior of $\mbox{\boldmath${F}$}$
because $\ddot g$ also vanishes at the zeroes of $g$
\cite{slih,PIQC}, and the potential term of (\ref{operator}) remains
analytic, but nevertheless the calculation of various quantities
associated with this operator becomes cumbersome due to the roots of
$g(\tau)$. Among such quantities is the functional determinant of
$\mbox{\boldmath${F}$}$ which determines the one-loop contribution
to the statistical sum of the CFT driven cosmology of \cite{PIQC}.
Its peculiarity is that the operator has an obvious zero mode which
is the function $g(\tau)$ itself,
    \begin{eqnarray}
    \mbox{\boldmath${F}$}g(\tau)=0,
    \end{eqnarray}
and the functional determinant of $\mbox{\boldmath${F}$}$ should, of
course, be understood as calculated on the subspace of its nonzero
modes. Thus, the focus of this paper is the calculation of such a
{\em restricted} functional determinant of (\ref{operator}), denoted
below by ${\rm Det_*}\mbox{\boldmath${F}$}$. This determinant is
calculated on the space of functions periodic on a compactified
range of the Euclidean time $\tau$ (forming a circle) with its zero
mode $g$ removed or gauged out.

There exist several different methods for restricted functional
determinants. When the whole spectrum of the operator is known this
is just the product of all non-zero eigenvalues. With the knowledge
of only the zero mode, one can use the regularization technique of
\cite{McK-Tarlie,Kirsten-McK,Kirsten-McK1} to extract the regulated
zero-mode eigenvalue from the determinant and subsequently take the
regularization off. Here we use another approach to the definition
of ${\rm Det_*}\mbox{\boldmath${F}$}$ dictated by the gauge-fixing
procedure for the path integral in cosmology \cite{PIQC}.

The zero mode of (\ref{operator}) arises in quantum cosmology as a
generator of the global gauge transformation of the gravitational
variable $\varphi$ (a canonically normalized perturbation of the
cosmological scale factor in the Friedmann-Robertson-Walker metric
\cite{PIQC}), $\Delta^\varepsilon\varphi(\tau)=\varepsilon g(\tau)$,
which is the residual symmetry of the action
    \begin{eqnarray}
    S[\,\varphi\,]=\frac12\oint
    d\tau\,\varphi(\tau)\mbox{\boldmath${F}$}
    \varphi(\tau),                          \label{action}
    \end{eqnarray}
remaining after gauge-fixing the local diffeomorphism invariance.
Here the integration runs over the full period of time compactified
to a circle according to the definition of the cosmological
statistical sum \cite{why}, and the variable $\varphi(\tau)$ as well
as the zero mode $g(\tau)$ are periodic on this circle. Therefore,
this symmetry is also subject to the Faddeev-Popov gauge fixing
procedure consisting of imposing the gauge on the integrated
(1-dimensional) field and inserting in  the path integral the the
relevant Faddeev-Popov factor.

This  gauge condition $\chi(\varphi)$ and the corresponding
Faddeev-Popov ghost factor $Q$ can be chosen in the form
    \begin{eqnarray}
    &&\chi=\oint d\tau\,g(\tau)\,\varphi(\tau),   \label{gauge0}\\
    &&\Delta^\varepsilon\chi=Q\varepsilon,\,\,\,
    Q=\oint d\tau\,g^2(\tau),
    \end{eqnarray}
and the restricted functional determinant of $\mbox{\boldmath${F}$}$
can be determined as the following Faddeev-Popov Gaussian functional
integral with the delta-function type gauge
    \begin{eqnarray}
    ({\rm Det}_*{\mbox{\boldmath${F}$}})^{-1/2}
    ={\rm const}\times\int D\varphi\;
    \delta\Big(\oint d\tau\, g\varphi\Big)\,Q\,
    \exp\left\{-\frac12\oint
    d\tau\,\varphi\,{\mbox{\boldmath${F}$}}\varphi\right\}.  \label{I}
    \end{eqnarray}
This definition is in fact independent of the choice of gauge by the
usual gauge independence mechanism for the Faddeev-Popov integral.
In particular, enforcing the gauge $\chi=0$ means that the field
$\varphi$ is functionally orthogonal to the zero mode $g(\tau)$ in a
trivial $L^2$ metric on a circle, and the above definition is
independent of the choice of this metric -- a possible function
weighting the integrand of (\ref{gauge0}).

In this paper we undertake the calculation of this path integral for
the restricted functional determinant of (\ref{operator}) as an
explicit functional of $g(\tau)$. The result will be obtained in
quadratures for a particular case when the function $g(\tau)$ has
one oscillation in the full range of the Euclidean time forming a
circle. Therefore, the function $g(\tau)$ has two zeroes at the
points labeled by $\tau_\pm$, $g(\tau_\pm)=0$, which will mark the
boundaries of the half period of the total time range,
$T=2(\tau_+-\tau_-)$. For brevity of the formalism we shift the
first root of $g(\tau)$ to zero, $\tau_-=0$, and let the coordinate
$\tau$ run in the total range $-\tau_+\leq\tau\leq\tau_+$ with the
points $\pm\tau_+$ identified. Then $g(\tau)$ is an odd function of
$\tau$ which is periodic with all its derivatives and has two first
degree zeros at antipodal points $\tau=\tau_-\equiv 0$ and
$\tau=\tau_+$ of this circle
    \begin{eqnarray}
    &&g(\tau)=-g(-\tau),\\
    &&g(\tau_\pm)=0,\,\,\,\dot g(\tau_\pm)
    \equiv \dot g_\pm\neq 0.                    \label{conditionsong}
    \end{eqnarray}
As mentioned above, in spite of singularity of $1/g$ at $\tau_\pm$
the operator (\ref{operator}) is everywhere regular (analytic) on
the circle, because $ \ddot g(\tau_\pm)=0$.

The final result of this paper -- the restricted functional
determinant of the operator (\ref{operator}) with the zero mode $g$
gauged out -- is determined by a special solution of the homogeneous
equation ${\mbox{\boldmath${F}$}}\,\varPsi=0$. This is a two-point
function $\Psi(\tau,\tau_*)$
    \begin{eqnarray}
    \varPsi(\tau,\tau_*)\equiv
    g(\tau)\int_{\tau_*}^{\tau}\frac{dy}{g^2(y)},\,\,\,\,\,
    \tau_-\equiv 0<\tau<\tau_+,\,\,\,\,\,
    \tau_-<\tau_*<\tau_+,                              \label{Psi}
    \end{eqnarray}
with some fixed point $\tau_*$ in the half period range of $\tau$.
The determinant of (\ref{operator}) in terms of this solution equals
   \begin{eqnarray}
    &&{\rm Det_*}\mbox{\boldmath${F}$}=
    {\rm const}\times
    \left|\;\varPsi_+\dot\varPsi_+
    -\varPsi_-\dot\varPsi_-\right|,           \label{det}\\
    &&\varPsi_\pm\equiv
    \varPsi(\tau_\pm,\tau_*),\,\,\,\,
    \dot\varPsi_\pm\equiv\dot\varPsi(\tau_\pm,\tau_*). \label{Psis}
    \end{eqnarray}

The main property of the function $\varPsi(\tau,\tau_*)$ is that it
is smoothly defined in the half-period range of $\tau$ and $\tau_*$,
where the integral (\ref{Psi}) is convergent because the roots of
$g(\tau)$ do not occur in the integration range. It cannot be
smoothly continued beyond the half-period
$\tau_-\leq\tau\leq\tau_+$, though its limits are well defined for
$\tau\to\tau_\pm\mp0$,
    \begin{eqnarray}
    \varPsi(\tau_\pm,\tau_*)=
    -\frac1{\dot g(\tau_\pm)}\equiv
    -\frac1{\dot g_\pm},                  \label{Psipm}
    \end{eqnarray}
because the factor $g(\tau)$ tending to zero compensates for the
divergence of the integral at $\tau\to\tau_\pm$. Moreover, because
of $\ddot g(\tau_\pm)=0$ the function $\varPsi(\tau,\tau_*)$ is
differentiable at $\tau\to\tau_\pm$, and all the quantities which
enter the algorithm (\ref{det}) are well defined. These properties
of $\varPsi(\tau,\tau_*)$ guarantee the independence of the obtained
result from an arbitrary choice of the point $\tau_*$, which can be
easily verified by using a simple relation
$d\dot\varPsi_\pm/d\tau_*=-\dot g_\pm/g^2(\tau_*)$.

\section{Variational expression for the determinant}
Representing the delta function of the gauge condition in (\ref{I})
via the integral over the Lagrangian multiplier $\pi$ we get the
Gaussian path integral over the periodic function $\varphi(\tau)$
and the numerical variable $\pi$ which are collectively denoted by
$\varPhi=(\varphi(\tau),\pi)$,
    \begin{eqnarray}
    &&({\rm Det}_*{\mbox{\boldmath${F}$}})^{-1/2}
    ={\rm const}\times Q\int D\varphi\,d\pi\,\exp\Big(-
    S_{\rm eff}[\,\varphi(\tau),\pi\,]\,\Big)
    ={\rm const}\times Q\,\Big({\rm Det}\,
    \mathbb{F}\Big)^{-1/2}.                         \label{I2}
    \end{eqnarray}
Here $S_{\rm eff}[\,\varphi(\tau),\pi\,]$ is the effective action of
these variables and $\mathbb{F}$ is the matrix valued Hessian of
this action with respect to $\varPhi$,
    \begin{eqnarray}
    &&S_{\rm eff}[\,\varphi(\tau),\pi\,]=
    \oint
    d\tau\,\Big(\,\frac12\,\varphi
    {\mbox{\boldmath${F}$}}\varphi-i\pi g\varphi\Big),\\
    &&\mathbb{F}=
    \frac{\delta^2S_{\rm eff}}
    {\delta\varPhi\, \delta\varPhi'}=
    \left[\,\begin{array}{cc} \;{\mbox{\boldmath${F}$}}\,
    \delta(\tau,\tau')&\,\,\, -i g(\tau)\,\\
    &\\
    -i g(\tau')&0\end{array}\,\right]     \label{matrixF}
    \end{eqnarray}
(note the position of time entries associated with the variables
$\varPhi=(\varphi(\tau),\pi)$ and $\varPhi'=(\varphi(\tau'),\pi)$).

A dependence of this determinant on $g$ can be found from its
functional variation with respect to $g(\tau)$. From (\ref{I2}) we
have
    \begin{eqnarray}
    &&\delta\ln\Big({\rm Det_*}\,
    \mbox{\boldmath${F}$}\Big)=-2\delta\ln Q+
    {\rm Tr}\,\Big(\delta\mathbb{F}\,\mathbb{G}\Big),   \label{var1}
    \end{eqnarray}
where $\mathbb{G}$ is the Green's function of $\mathbb{F}$,
$\mathbb{F}\,\mathbb{G}=\mathbb{I}$. The block structure of this
matrix Green's function has the form
    \begin{eqnarray}
    &&\mathbb{G}=
    \left[\,\,\begin{array}{cc}G(\tau,\tau')\,&\,\,\,
    {\displaystyle \frac{\textstyle i g(\tau)}Q}\,\\
    &\\
    {\displaystyle \frac{\textstyle i g(\tau')}Q}
    &0\end{array}
    \,\right],                                    \label{matrixG}
    \end{eqnarray}
where the Green's function $G(\tau,\tau')$ in the diagonal block
satisfies the system of equations
    \begin{eqnarray}
    &&{\mbox{\boldmath$F$}}\,
    G(\tau,\tau')=\delta(\tau,\tau')
    -\frac{g(\tau)\,g(\tau')}Q,            \label{Gequation}\\
    &&\oint d\tau\,g(\tau)\,G(\tau,\tau')
    =0,                                           \label{Ggauge}
    \end{eqnarray}
which uniquely fix it. The second equation imposes the needed gauge,
whereas the right hand side of the first equation implies that
$G(\tau,\tau')$ is the inverse of the operator $F$ on the subspace
orthogonal to its zero mode. In what follows it will be useful to
express $G(\tau,\tau')$ in terms of another auxiliary Green's
function $\tilde G(\tau,\tau')$. It arises as follows. From
(\ref{Gequation})-(\ref{G}) it follows that the Green's function
$G(\tau,\tau')$ gives the solution
    \begin{eqnarray}
    \tilde\varphi(\tau)=\oint d\tau'\,
    G(\tau,\tau')\,
    J(\tau')                      \label{tildevarphi0}
    \end{eqnarray}
to the following problem
    \begin{eqnarray}
    &&{\mbox{\boldmath$F$}}
    \tilde\varphi(\tau)=\tilde J(\tau),    \label{equation}\\
    &&\oint d\tau\, g\tilde\varphi=0,          \label{gauge}\\
    &&\tilde J(\tau)\equiv J(\tau)-
    \frac{g(\tau)}Q\oint d\tau_1\,g(\tau_1)J(\tau_1)
    ,\,\,\,\,
    \oint d\tau\,g\tilde J\equiv 0,         \label{tildeJ}
    \end{eqnarray}
with the modified source $\tilde J(\tau)$ which is functionally
orthogonal to the zero mode $g(\tau)$ -- the property that
guarantees the existence of the solution of Eq.(\ref{equation})
whose left hand side is also functionally orthogonal to $g$ in view
of integration by parts in $\oint
d\tau\,g\,({\mbox{\boldmath$F$}}\tilde\varphi)=0$. If we denote by
$\tilde G(\tau,\tau')$ an auxiliary Green's function which solves
this last problem in the form
    \begin{eqnarray}
    \tilde\varphi(\tau)=\oint d\tau\,
    \tilde G(\tau,\tau')\,\tilde J(\tau'),   \label{tildevarphi}
    \end{eqnarray}
then the comparison with (\ref{tildevarphi0}) shows the relation
between the two Green's functions via the projection of $\tilde
G(\tau,\tau')$ with respect to the second argument onto the subspace
orthogonal to the zero mode
    \begin{eqnarray}
    &&G(\tau,\tau')=
    \tilde G(\tau,\tau')-
    \oint d\tau_1\,\tilde G(\tau,\tau_1)\,
    \frac{g(\tau_1)\,g(\tau')}Q\,.          \label{G}
    \end{eqnarray}

The introduction of $\tilde G(\tau,\tau')$ is justified by a number
of simplifications which one has when finding the solution
$\tilde\varphi(\tau)$ in terms of $\tilde J(\tau)$ rather than in
terms of the original source. This remains true even despite the
fact that this auxiliary Green's function is not unique because of
the freedom in the transformation $\tilde G(\tau,\tau')\to \tilde
G(\tau,\tau')+\epsilon(\tau)\,g(\tau')$ preserving
(\ref{tildevarphi}). This freedom results in the ambiguity of the
equation for $\tilde G(\tau,\tau')$,
    \begin{eqnarray}
    {\mbox{\boldmath$F$}}\tilde
    G(\tau,\tau')=\delta(\tau,\tau')
    +\omega(\tau)\,g(\tau'),                    \label{omega}
    \end{eqnarray}
with some function $\omega(\tau)$ which is transformed as
$\omega(\tau)\to \omega(\tau)+\mbox{\boldmath$F$}\epsilon(\tau)$.
The function $\omega(\tau)$ can be rather general and is only
subject to the condition $\oint d\tau g\omega=-1$ which follows from
integrating the above equation with $g(\tau)$ -- the zero mode of
$\mbox{\boldmath$F$}$.\footnote{This does not necessarily imply that
$\omega(\tau)=-g(\tau)/\int dy\,g^2(y)$, as one would expect from
the functional orthogonality of $g(\tau)$ and $G(\tau,\tau')$. This
will be confirmed by the calculation of a specific $\omega(\tau)$
below.} This ambiguity, of course, goes away in the Green's function
(\ref{G}) due to the projection operation.

Now we can express the variation (\ref{var1}) in terms of the above
Green's functions. Using (\ref{matrixF}) and (\ref{matrixG}) in
(\ref{var1}) we have
    \begin{eqnarray}
    &&\delta\ln\Big({\rm Det_*}\,
    \mbox{\boldmath$F$}\Big)=
    \oint d\tau\,\delta\mbox{\boldmath$F$}\,
    G(\tau,\tau')\Big|_{\,\tau'=\tau}-\delta\ln Q.
    \end{eqnarray}
Also substituting (\ref{G}) in the first term and integrating by
parts the action of the operator $\delta\mbox{\boldmath$F$}$ we have
    \begin{eqnarray}
    &&\oint d\tau\,\delta\mbox{\boldmath$F$}\,G\,
    \Big|_{\tau'=\tau}=\oint d\tau\,
    \delta\mbox{\boldmath$F$}\,
    \tilde G\,\Big|_{\,\tau'=\tau}-\frac1Q\oint
    d\tau\,dy\,\big(\delta\mbox{\boldmath$F$}
    \,g(\tau)\big)\,\tilde
    G(\tau,y)\,g(y),
    \end{eqnarray}
whence on account of the obvious variational relation
$\delta\mbox{\boldmath$F$}\,g=-\mbox{\boldmath$F$}\,\delta g$ and
again integrating by parts we have
    \begin{eqnarray}
    &&\oint d\tau\,\delta\mbox{\boldmath$F$}\,G\,\Big|_{\,\tau'=\tau}
    =\oint d\tau\,\delta\mbox{\boldmath$F$}\,
    \tilde G\,\Big|_{\,\tau'=\tau}+\frac12\,
    \delta\ln Q+\oint d\tau\,\delta g(\tau)\,\omega(\tau),
    \end{eqnarray}
where we have used (\ref{omega}). Therefore finally
    \begin{eqnarray}
    &&\delta\ln\Big({\rm Det_*}\,\mbox{\boldmath$F$}\Big)=
    \oint d\tau\,\delta\mbox{\boldmath$F$}\,
    \tilde G(\tau,\tau')\,\Big|_{\,\tau'=\tau}
    -\frac12\,\delta\ln Q
    +\oint d\tau\,\delta g(\tau)\,\omega(\tau).      \label{Detprime}
    \end{eqnarray}
One can check that this expression is also invariant with respect to
the $\epsilon$-transformation of $\tilde G(\tau,\tau')$ and
$\omega(\tau)$ of the above type. Below we explicitly find $\tilde
G(\tau,\tau')$ along with its $\omega(\tau)$ and calculate the above
variation.

\section{The periodic boundary conditions problem
and its Green's function}

We will look for the periodic solution of the problem
(\ref{equation})-(\ref{gauge}) as a linear combination of the
partial solution of the inhomogeneous equation (\ref{equation}) and
two solutions of the homogeneous equation -- zero mode $g(\tau)$ and
the function $\varPsi(\tau,\tau_*)$ defined in Introduction by
Eq.(\ref{Psi}). A partial solution of the inhomogeneous equation
(\ref{equation}) reads
    \begin{eqnarray}
    &&\varPhi(\tau)=-g(\tau)\int_{0}^\tau
    \frac{dy}{g^2(y)}\int_{0}^{y} d\tau'\,g\tilde
    J(\tau')=-\int_{0}^\tau d\tau'\,\varPsi(\tau,\tau')\,
    g(\tau')\tilde J(\tau')\nonumber\\
    &&\qquad\qquad
    =-\oint d\tau'
    \Big[\;\theta(\tau-\tau')\,\theta(\tau')-
    \theta(\tau'-\tau)\,\theta(-\tau')\,\Big]\,
    \varPsi(\tau,\tau')\,
    g(\tau')\tilde J(\tau').               \label{Phi}
    \end{eqnarray}
It is defined for all $-\tau_+\leq\tau\leq\tau_+$ and satisfies at
$\tau=\tau_-$ the initial conditions
    \begin{eqnarray}
    \varPhi(0)=0,\,\,\,\,\,\dot\varPhi(0)=0.   \label{Phi(0)}
    \end{eqnarray}
However it is not periodic, because
$\varPhi(-\tau_+)=\varPhi(\tau_+)$, but
$\dot\varPhi(-\tau_+)\neq\dot\varPhi(\tau_+)$. Indeed, in view of
(\ref{Psipm}) and (\ref{tildeJ})
    \begin{eqnarray}
    &&\varPhi(\tau_+)-\varPhi(-\tau_+)=\frac1{\dot
    g_+}\left(\,\int_{0}^{\tau_+}
    -\int_{0}^{-\tau_+}\right)
    d\tau \,g\tilde J=
    \frac1{\dot g_+}\oint
    d\tau\,g\tilde J(\tau)=0,                    \label{Phipm}\\
    &&\dot\varPhi(\tau_+)-\dot\varPhi(-\tau_+)=
    -\int_{0}^{\tau_+} dy\,\dot\varPsi(\tau_+,y)\,
    g(y)\tilde J(y)
    +\int_{0}^{-\tau_+} dy\,\dot\varPsi(-\tau_+,y)\,
    g(y)\tilde J(y)\nonumber\\
    &&\qquad\qquad\qquad\qquad\quad
    =-\oint dy
    \left[\;\dot\varPsi(\tau_+,y)\,\theta(y)
    -\dot\varPsi(\tau_+,-y)\,
    \theta(-y)\,\right]
    g(y)\tilde J(y)\neq 0.                   \label{dotPhipm}
    \end{eqnarray}
Here we used the fact that in view of the asymmetry of $g(\tau)$,
$g(-\tau)=-g(\tau)$, $\varPsi(-\tau,y)=\varPsi(\tau,-y)$ and
$\dot\varPsi(-\tau,y)=-\dot\varPsi(\tau,-y)$.

Thus, we shall look for the solution in question as a linear
combination of $\Phi(\tau)$ and two basis functions of
$\mbox{\boldmath$F$}$ -- the periodic function $g(\tau)$ and the
non-periodic $\varPsi(\tau,\tau')$ with some $\tau'=\tau_*>0$. For
negative $\tau$ the role of this second basis function will be
played by $\varPsi(-\tau,\tau_*)$, because
$\mbox{\boldmath$F$}\,\varPsi(-\tau,\tau_*)=0$ in view of the odd
nature of $g(\tau)$ in (\ref{operator}). Thus the solution has the
following {\em piecewise smooth} form
    \begin{eqnarray}
    &&\tilde\varphi(\tau)=\varPhi(\tau)
    +C_+\,\varPsi(\tau,\tau_*)\,\theta(\tau)
    +C_-\,\varPsi(-\tau,\tau_*)\,\theta(-\tau)\nonumber\\
    &&\qquad\qquad\qquad\qquad\qquad\qquad
    +D_+\,g(\tau)\,\theta(\tau)
    +D_-\,g(\tau)\,\theta(-\tau),            \label{ansatz}
    \end{eqnarray}
where all four coefficients should be determined from the periodic
boundary conditions at $\tau=\pm\tau_+$,
    \begin{eqnarray}
    &&\tilde\varphi(\tau_+)=\tilde\varphi(-\tau_+),  \label{periodicbc1}\\
    &&\dot{\tilde\varphi}(\tau_+)=
    \dot{\tilde\varphi}(-\tau_+),                    \label{periodicbc2}.
    \end{eqnarray}
the gauge condition (\ref{gauge}) and also the conditions of
smoothness at $\tau=\tau_-\equiv 0$
    \begin{eqnarray}
    &&\tilde\varphi(+0)=\tilde\varphi(-0),  \label{smoothness1}\\
    &&\dot{\tilde\varphi}(0)=
    \dot{\tilde\varphi}(-0).                    \label{smoothness2}
    \end{eqnarray}

In view of $g(0)=0$ together with (\ref{Phi(0)}) and the fact that
$\varPsi(\tau,\tau_*)$ in (\ref{ansatz}) is continued to negative
values of $\tau$ as an even function, the condition
(\ref{smoothness1}) implies that $C_+=C_-$. The second of smoothness
conditions reads $C_+\dot\varPsi_-+D_+\dot
g_-=-C_-\dot\varPsi_-+D_+\dot g_-$,
$\dot\varPsi_-\equiv\dot\varPsi(0,\tau_*)$, and results in
    \begin{eqnarray}
    2C_+\,\dot\varPsi_-=-(D_+-D_-)\,\dot g_-.
    \end{eqnarray}
The first of periodicity conditions (\ref{periodicbc1}) is
identically satisfied because $g(\pm\tau_+)=0$, $C_-=C_+$ and
(\ref{Phipm}), while the second condition (\ref{periodicbc2}) reads
$\dot\varPhi(\tau_+)+C_+\,\dot\varPsi_++D_+\,\dot
g_+=\dot\varPhi(-\tau_+)-C_-\,\dot\varPsi_++D_-\,\dot g(-\tau_+)$,
where $\dot\varPsi_+\equiv\dot\varPsi(\tau_+,\tau_*)$, or
    \begin{eqnarray}
    \dot\varPhi(\tau_+)-\dot\varPhi(-\tau_+)
    +2C_+\,\dot\varPsi_++(D_+-D_-)\,\dot g_+=0.
    \end{eqnarray}
Therefore in view of (\ref{Psipm})
    \begin{eqnarray}
    &&C_\pm=-\frac12\,
    \frac{\varPsi_+}{\varPsi_+\dot\varPsi_+
    -\varPsi_-\dot\varPsi_-}\,
    \big[\,\dot\varPhi(\tau_+)
    -\dot\varPhi(-\tau_+)\big],          \label{Cpm}\\
    &&D_+ -D_- =-\varPsi_+\,
    \frac{\varPsi_-\dot\varPsi_-}
    {\varPsi_+\dot\varPsi_+
    -\varPsi_-\dot\varPsi_-}\,
    \big[\,\dot\varPhi(\tau_+)
    -\dot\varPhi(-\tau_+)\big].         \label{D-D}
    \end{eqnarray}

In view of the odd nature of $g(\tau)$, $g(-\tau)=-g(\tau)$, the
$C_\pm$ coefficients do not contribute to the last equation on the
coefficients of (\ref{ansatz}) --- the gauge condition
(\ref{gauge}). This gauge condition leads to
    \begin{eqnarray}
    \frac{D_++D_-}2=-\frac1Q \oint
    d\tau\,g\,\varPhi.                  \label{halfofD+D}
    \end{eqnarray}

Thus, all coefficients are uniquely determined by the source $\tilde
J$. Since $\varPhi$, $C_\pm$ and $D_\pm$ are all linear in this
source the Green's function $\tilde G(\tau,\tau')$ can be read off
(\ref{ansatz}) as
    \begin{eqnarray}
    \tilde G(\tau,\tau')
    =\frac{\delta\tilde\varphi(\tau)}
    {\delta\tilde J(\tau')},                 \label{vartildeG}
    \end{eqnarray}
where of course the functional derivative should not be understood
literally because the modified source is functionally restricted by
the condition of orthogonality to the zero mode (\ref{tildeJ}).
Rather, this expression should be regarded as a coefficient of
$\tilde J(\tau')$ in Eq.(\ref{tildevarphi}).

Now we have to determine the function $\omega$ in the equation
(\ref{omega}) for this Green's function. From (\ref{ansatz}) it
follows that $\tilde G(\tau,\tau')$ is a priori smooth and satisfies
the homogeneous equation $\mbox{\boldmath$F$}\tilde G(\tau,\tau')=0$
for all $\tau\neq \tau',0,\tau_+$. At $\tau=\tau'$ it is continues
but has a jump in the first order derivative which contributes the
delta $\delta(\tau,\tau')$ to the right hand side of (\ref{omega}).
The continuity of $\tilde\varphi(\tau)$ and its derivative at
$\tau=\tau_-$ and $\tau=\pm\tau_+$ could have implied complete
smoothness of $\tilde G(\tau,\tau')$ at these points if the
expression (\ref{tildevarphi}) for $\tilde\varphi(\tau)$ would hold
for an arbitrary source $\tilde J$. However this source is not
arbitrary -- it is functionally orthogonal to $g(\tau)$, $\int
d\tau\,g\tilde J=0$, and the continuity of $\tilde\varphi(\tau)$ at
$\pm\tau_+$, (\ref{periodicbc1}), holds only in virtue of this
orthogonality (cf. Eq.(\ref{Phipm})). Therefore the Green's function
$\tilde G(\tau,\tau')$ can have a discontinuity at this point
proportional to $g(\tau')$, which disappears in $\tilde\varphi$
after being integrated with $\tilde J(\tau')$. This discontinuity is
contributed by the $\varPhi$-part of $\tilde\varphi$ or by the jump
of the kernel in the last line of the equation (\ref{Phi})
    \begin{eqnarray}
    &&\tilde G(\tau_+,\tau')-\tilde G(-\tau_+,\tau')\nonumber\\
    &&\qquad\qquad\quad
    =-\Big[\;\theta(\tau-\tau')\,\theta(\tau')-
    \theta(\tau'-\tau)\,\theta(-\tau')\,\Big]\,
    \varPsi(\tau,\tau')\,
    g(\tau')\,\Big|_{\,\tau=-\tau_+}^{\,\tau=\tau_+}
    =\frac1{\dot g_+}\,g(\tau').                   \label{jump}
    \end{eqnarray}
On the contrary, the first order derivative of $\tilde
G(\tau,\tau')$ at $\pm\tau_+$ is continuous, because the continuity
of $\dot{\tilde\varphi}$ at this points is enforced irrespective of
the choice of the source.

The discontinuity (\ref{jump}) contributes the second term on the
right hand side of the equation (\ref{omega}) for $\tilde
G(\tau,\tau')$
    \begin{eqnarray}
    \mbox{\boldmath$F$}\tilde G(\tau,\tau')=\delta(\tau,\tau')+
    \frac1{\dot g_+}\,
    \dot\delta(\tau,\tau_+)\,g(\tau'),   \label{equationfortildeG}
    \end{eqnarray}
when treated as a generalized function in the vicinity of the point
$\tau=\pm\tau_+$. Indeed, since the interval $-\tau_+<\tau<\tau_+$
forms a circle with the points $\pm\tau_+$ identified, a small
negative vicinity of the point $\tau_+$ becomes adjacent to a small
positive vicinity of $-\tau_+$. Shifting for brevity this point to
zero, $\tau_+\to\tau=0$, we then have such a situation. The full
Greens function has a form $G=G_1\theta(\tau)+G_2\theta(-\tau)$ with
two different functions $G_1(\tau)$ and $G_2(\tau)$ having different
limits at zero, $G_1(0)\neq G_2(0)$, but the same derivatives, $\dot
G_1(0)=\dot G_2(0)$, and satisfying one and the same equation
$\mbox{\boldmath$F$}G_{1,2}=\delta(\tau,\tau')$. Then
    \begin{eqnarray}
    &&\mbox{\boldmath$F$}G=
    (\mbox{\boldmath$F$}G_1)\theta(\tau)
    +(\mbox{\boldmath$F$}G_2)\theta(-\tau)
    -2[\dot G_1(0)-\dot G_2(0)]\,\delta(\tau)\nonumber\\
    &&\qquad\quad-[G_1(0)-G_2(0)]\,\dot\delta(\tau)
    =\delta(\tau,\tau')-[G_1(0)-G_2(0)]\,
    \dot\delta(\tau),
    \end{eqnarray}
which gives (\ref{equationfortildeG}). Therefore, $\omega(\tau)$ in
eq.(\ref{omega}) equals
    \begin{eqnarray}
    \omega(\tau)=
    \frac1{\dot g_+}\,\dot\delta(\tau,\tau_+),       \label{omega1}
    \end{eqnarray}
and of course satisfies the consistency condition $\oint
d\tau\,\omega(\tau)\,g(\tau)=-1$.

\section{The variation of the determinant}
In view of (\ref{vartildeG}) and (\ref{ansatz}) and the fact that
$\delta\mbox{\boldmath$F$}=\delta(\ddot g/g)$ one can write
    \begin{eqnarray}
    &&\oint d\tau\,\delta\mbox{\boldmath$F$}\,
    \tilde G(\tau,\tau')\Big|_{\tau'=\tau}
    =\oint d\tau\,\delta\!\left(\frac{\ddot g}g\right)\,
    \frac{\delta\varPhi(\tau)}
    {\delta\tilde J(\tau)}\nonumber\\
    &&\qquad\qquad\qquad
    +\oint d\tau\,
    \delta\!\left(\frac{\ddot g}g\right)\,
    \left(\,\varPsi(\tau,\tau_*)\,
    \frac{\delta C_+}{\delta\tilde J(\tau)}\,\theta(\tau)
    +\varPsi(-\tau,\tau_*)\,\frac{\delta C_-}{\delta\tilde
    J(\tau)}\,\theta(-\tau)\right)\nonumber\\
    &&\qquad\qquad\qquad+\oint d\tau\,
    \delta\!\left(\frac{\ddot g}g\right)\,g(\tau)
    \left(\,\frac{\delta D_+}
    {\delta\tilde J(\tau)}\,\theta(\tau)
    +\frac{\delta D_-}{\delta\tilde J(\tau)}\,
    \theta(-\tau)\right).                             \label{C+D}
    \end{eqnarray}
From (\ref{Phi}) it follows that $\delta\varPhi(\tau)/\delta\tilde
J(\tau)\propto\varPsi(\tau,\tau)=0$, and the first term here
vanishes.

Since $C_-=C_+$ the second term in (\ref{C+D}) is just the value of
$C_+=C_+[\,\tilde J(\tau)\,]$ under a special choice of the source
$\tilde J$,
    \begin{eqnarray}
    &&\oint d\tau\,\delta\!\left(\frac{\ddot g}g\right)\,
    \left(\,\varPsi(\tau,\tau_*)\,
    \frac{\delta C_+}{\delta\tilde J(\tau)}\,\theta(\tau)
    +\varPsi(-\tau,\tau_*)\,\frac{\delta C_-}{\delta\tilde
    J(\tau)}\,\theta(-\tau)\right)\nonumber\\
    &&\qquad\qquad\qquad\qquad\qquad
    =C_+\Big|_{\,\textstyle\tilde J(\tau)=
    \delta(\ddot g/g)
    \big(\varPsi(\tau,\tau_*)\theta(\tau)
    +\varPsi(-\tau,\tau_*)\theta(-\tau)\big)}. \label{C00}
    \end{eqnarray}
This expression is calculated in Appendix A and reads
    \begin{eqnarray}
    &&C_+\Big|_{\,\textstyle\tilde J(\tau)=
    \delta(\ddot g/g)
    \big(\varPsi(\tau,\tau_*)\theta(\tau)
    +\varPsi(-\tau,\tau_*)\theta(-\tau)\big)}\nonumber\\
    &&\qquad\qquad\quad
    =-\frac1{\varPsi_+\dot\varPsi_+
    -\varPsi_-\dot\varPsi_-}\,\Big(\varPsi_-\,\delta\dot\varPsi_-
    -\delta\varPsi_-\,\dot\varPsi_-
    -\varPsi_+\,\delta\dot\varPsi_+
    +\frac{\varPsi_+\dot\varPsi_+}{\varPsi_-}\,
    \delta\varPsi_-\Big).                 \label{C}
    \end{eqnarray}
D-terms of (\ref{C+D}) can also be reorganized as
    \begin{eqnarray}
    &&\oint d\tau\,
    \delta\!\left(\frac{\ddot g}g\right)\,g(\tau)
    \left(\,\frac{\delta D_+}{\delta\tilde J(\tau)}\,\theta(\tau)
    +\frac{\delta D_-}{\delta\tilde
    J(\tau)}\,\theta(-\tau)\right)\nonumber\\
    &&\qquad\qquad\qquad
    =\left.\frac{D_++D_-}2\,
    \right|_{\,\textstyle\tilde J=\delta(\ddot g/g)\,g(\tau)}
    +\left.\frac{D_+-D_-}2\,
    \right|_{\,\textstyle\tilde J
    =\delta(\ddot g/g)\,g\,\varepsilon(\tau)},   \label{Dgeneral}
    \end{eqnarray}
where $\varepsilon(\tau)=\theta(\tau)-\theta(-\tau)$. The
calculations of Appendix C give
    \begin{eqnarray}
    &&\left.\frac{D_++D_-}2\,
    \right|_{\,\textstyle\tilde J=\delta(\ddot g/g)\,g(\tau)}
    =\frac12\,\delta\ln Q+
    \frac{\delta\varPsi_-}{\varPsi_-},               \label{D1}\\
    &&\left.\frac{D_+-D_-}2\,
    \right|_{\,\textstyle\tilde J
    =\delta(\ddot g/g)\,g\,
    \varepsilon(\tau)}
    =\frac{\varPsi_-\dot\varPsi_-}
    {\varPsi_+\dot\varPsi_+
    -\varPsi_-\dot\varPsi_-}\,
    \left(\frac{\delta \varPsi_+}{\varPsi_+}-
    \frac{\delta \varPsi_-}{\varPsi_-}\right).  \label{D2}
    \end{eqnarray}
Collecting (\ref{C00})-(\ref{C}), (\ref{D1}) and (\ref{D2}) in
(\ref{C+D}) we have
    \begin{eqnarray}
    &&\oint d\tau\,\delta\mbox{\boldmath${F}$}\,
    \tilde G(\tau,\tau')\,\Big|_{\,\tau'=\tau}
    =\delta\ln\left|\,\frac{\varPsi_+\dot\varPsi_+
    -\varPsi_-\dot\varPsi_-}{\varPsi_+}\right|
    +\frac12\,\delta\ln Q.
    \end{eqnarray}

After substituting this expression in (\ref{Detprime}) with
$\omega(\tau)$ given by (\ref{omega1}) one can see that the
contributions of the Faddeev-Popov factor $Q$ completely cancel out.
Therefore, as expected gauging out the zero mode of
$\mbox{\boldmath${F}$}$ is gauge independent, and we get
    \begin{eqnarray}
    &&\delta\ln\Big({\rm Det_*}\mbox{\boldmath${F}$}\Big)=
    \delta\ln\left|\,\frac{\varPsi_+\dot\varPsi_+
    -\varPsi_-\dot\varPsi_-}{\varPsi_+}\right|
    +\frac1{\dot g_+}\oint d\tau\,\delta
    g(\tau)\,\dot\delta(\tau,\tau_+)\nonumber\\
    &&\qquad\qquad\qquad=\delta\ln\left|\,\frac{\varPsi_+\dot\varPsi_+
    -\varPsi_-\dot\varPsi_-}{\varPsi_+}\right|
    -\frac{\delta\dot g_+}{\dot g_+}=
    \delta\ln\left|\;\varPsi_+\dot\varPsi_+
    -\varPsi_-\dot\varPsi_-\right|,
    \end{eqnarray}
which finally proves (\ref{det}).

\section{The case of the second zero mode}
In the special case when the basic quantity of Eq.(\ref{det})
degenerates to zero,
    \begin{eqnarray}
    \varPsi_+\dot\varPsi_+
    -\varPsi_-\dot\varPsi_-=0,  \label{degeneration}
    \end{eqnarray}
the coefficients $C_\pm$ and $D_\pm$ in the solution (\ref{ansatz}),
defined by (\ref{Cpm})-(\ref{halfofD+D}), become singular and also
the restricted functional determinant vanishes
    \begin{eqnarray}
    {\rm Det_*}\mbox{\boldmath${F}$}=0.   \label{zerodet}
    \end{eqnarray}
All this points out to the fact that the operator
$\mbox{\boldmath${F}$}$ has an additional zero mode different from
$g(\tau)$. This mode is not gauged out in the path integral, and the
corresponding functional determinant vanishes due to its zero
eigenvalue.

This zero mode can be constructed as follows. Take the basis
function $\varPsi(\tau,\tau_*)$ of the operator, defined by Eq.
(\ref{Psi}) on the half period $\tau_+\geq\tau\geq 0$. It satisfies
initial conditions $(\varPsi_-,\dot\varPsi_-)$ at $\tau_-=0$. Now,
smoothly continue this function to the negative half period of
$\tau$ by evolving these initial conditions by the equation of
motion backward in time. To be more precise, this function denoted
below as $\mbox{\boldmath${\varPsi}$}(\tau)$ will be a solution of
the Cauchy problem
    \begin{eqnarray}
    &&\mbox{\boldmath${F}$}
    \mbox{\boldmath${\varPsi}$}(\tau)=0,\,\,\,\,-\tau_+<\tau<\tau_+,\\
    &&\mbox{\boldmath${\varPsi}$}(0)=\varPsi_-,\,\,\,\,
    \dot{\mbox{\boldmath${\varPsi}$}}(0)=\dot\varPsi_-,
    \end{eqnarray}
with the Cauchy data surface in the middle of the time range at
$\tau_-=0$. For a positive $\tau$ it coincides with
$\varPsi(\tau,\tau_*)$, whereas for $\tau<0$ this is a linear
combination of $\varPsi(\,|\,\tau\,|,\tau_*)$ and $g(\tau)$ with the
coefficients following from the initial data at $\tau_-$,
    \begin{eqnarray}
    \mbox{\boldmath${\varPsi}$}(\tau)
    =\varPsi(\,|\,\tau\,|,\tau_*)
    -2\varPsi_-\dot\varPsi_-\,
    \theta(-\tau)\,g(\tau).       \label{second0mode}
    \end{eqnarray}
Generically this function is not periodic on a circle
$-\tau_+\leq\tau\leq\tau_+$ with the points $\pm\tau_+$ identified,
because $\mbox{\boldmath${\varPsi}$}(+\tau_+)
=\mbox{\boldmath${\varPsi}$}(-\tau_+)$ in view of $g(\pm\tau_+)=0$,
but
    \begin{eqnarray}
    &&\dot{\mbox{\boldmath${\varPsi}$}}(-\tau_+)
    =\dot{\mbox{\boldmath${\varPsi}$}}(\tau_+)
    -\frac2{\varPsi_+}\,
    \big(\,\varPsi_+\dot\varPsi_+
    -\varPsi_-\dot\varPsi_-\big).
    \end{eqnarray}
However, for a special case of (\ref{degeneration}) the derivatives
also match at $\tau=\pm\tau_+$, and (\ref{second0mode}) becomes
completely smooth and periodic and forms the second zero mode of
$\mbox{\boldmath${F}$}$. This explains the vanishing value of the
restricted determinant and, in fact, confirms the validity of its
algorithm (\ref{det}).

A simple example of the operator \ref{operator}) with two zero modes
is given by the case of its constant potential term $\ddot
g/g=-w^2={\rm const}$, corresponding to a harmonic function
    \begin{eqnarray}
    g(\tau)=d\,\sin(w\tau), \,\,\,\,
    -\pi/w\leq\tau\leq\pi/w.  \label{sine}
    \end{eqnarray}
In this case the function $\varPsi(\tau,\tau_*)$ is exactly
calculable,
    \begin{eqnarray}
    \varPsi(\tau,\tau_*)=
    \frac{\sin\big(w(\tau-\tau_*)\big)}{d\,\sin(w\tau_*)},
    \end{eqnarray}
and it exactly satisfies the relation (\ref{degeneration}). This
function by itself is periodic on a circle and comprises the second
linear independent basis function of the equation
$\ddot\varPsi+w^2\varPsi=0$, complementary to (\ref{sine}).
Therefore, it can be taken as the second zero mode which would
differ from (\ref{second0mode}) only by a linear recombination with
$g(\tau)$.

The situation with the second zero mode of $\mbox{\boldmath${F}$}$
serves as a check of consistency of the obtained algorithm
(\ref{det}). Otherwise, its role should not be overestimated. In
particular, this mode does not have to be gauged out in the path
integral for the statistical sum in cosmology \cite{PIQC}. Unlike
$g(\tau)$ this second zero mode cannot be associated with the
generator of the invariance transformation for the whole
cosmological action. It leaves invariant the action of the
$\varphi$-variable (\ref{action}), but the full cosmological action
contains also the part with the global (modular) degree of freedom
-- the proper time period of the Euclidean FRW metric -- which is
not invariant under the transformations associated with the second
zero mode \cite{PIQC}. Consequently, the one-loop preexponential
factor of the statistical sum remains well-defined also in the limit
of (\ref{degeneration}), because the determinant (\ref{I2}) enters
this prefactor in a way more complicated than a simple overall
factor \cite{PIQC}. This is the result of the additional integration
over the global modular variable of the FRW metric. In applications
of the above technique to the microcanonical ensemble in cosmology
the second zero mode of $\mbox{\boldmath${F}$}$ arises for a special
subset of the cosmological instantons. As shown in \cite{oneloop1},
it leads in view of (\ref{degeneration})-(\ref{zerodet}) to a
certain suppression of loop corrections in the cosmological
statistical sum, but does not completely nullify them.

\section{Conclusions}
Though very involved, the above calculation of the functional
determinant of the operator (\ref{operator}) with its zero mode
gauged out leads to a rather concise answer (\ref{det}) in terms of
the special basis function of this operator (\ref{Psi}) and the zero
mode itself. The determinant ${\rm Det_*}\mbox{\boldmath${F}$}$ as
well as the solution of the periodic boundary conditions problem
with this operator enter the construction of the statistical sum in
quantum cosmology \cite{PIQC}. In particular, the form of
(\ref{det}) is crucial for the partial cancelation of contributions
to its one-loop preexponential factor. This cancelation is
associated with the absence of the local degrees of freedom in the
Friedmann-Robertson-Walker sector of cosmology. The remaining part
is due to the global degree of freedom of the FRW metric, and it is
also determined by the contribution of (\ref{det}).

Important limitation of the obtained result is, however, that it
includes only the case of two roots of the zero mode $g(\tau)$ in
the full period of the Euclidean time, whereas the path integral in
cosmology driven by a conformal field theory suggests cosmological
instantons with an arbitrary number of oscillations of the
cosmological scale factor, corresponding to numerous roots of the
oscillating zero mode $g$. For the number of these oscillations
tending to infinity this CFT driven cosmology approaches a new
quantum gravity scale -- the maximum possible value of the
cosmological constant \cite{slih,why} -- where the physics and, in
particular, the effects of the quantum prefactor become very
interesting and important. Thus the extension of the above technique
for ${\rm Det_*}\mbox{\boldmath${F}$}$ to an arbitrary number of
roots of the zero mode of $\mbox{\boldmath${F}$}$ becomes important,
as this extension might be relevant to the cosmological constant
problem.

The invariant definition of the restricted functional determinant
with the zero mode gauged out by the Faddeev-Popov method is likely
to be equivalent to the regularization technique of
\cite{McK-Tarlie,Kirsten-McK,Kirsten-McK1} for similar operators
with a zero eigenvalue. When combined with a monodromy method of
\cite{Forman} this technique might allow us to make this extension
manageable, which we hope to attain in a foreseeable future.

\section*{Acknowledgements}
A.B. wishes to express his gratitude to G.Dvali for hospitality at
the Physics Department of the Ludwig-Maximilians University in
Munich where this work was supported in part by the Humboldt
Foundation. The work of A.B. was also supported by the RFBR grant No
11-02-00512. The work of A.K. was supported by the RFBR grant No 08-02-00923.

\appendix
\renewcommand{\thesection}{Appendix \Alph{section}.}
\renewcommand{\theequation}{\Alph{section}.\arabic{equation}}

\section{The contribution of the C-terms}
The contribution of the C-terms (\ref{C00}) can be obtained by using
(\ref{dotPhipm}) and (\ref{Cpm})
    \begin{eqnarray}
    &&C_+\Big|_{\,\textstyle\tilde J(\tau)=
    \delta(\ddot g/g)\big(\varPsi(\tau,\tau_*)\theta(\tau)
    +\varPsi(-\tau,\tau_*)\theta(-\tau)\big)}\nonumber\\
    &&\qquad\qquad\qquad=\frac{\varPsi_+}
    {\varPsi_+\dot\varPsi_+-\varPsi_-\dot\varPsi_-}\,
    \int_{\tau_-}^{\tau_+} d\tau \left(\delta\ddot
    g\;\varPsi(\tau,\tau_*)-\frac{\ddot
    g}g\,\varPsi(\tau,\tau_*)\,\delta g\right)\dot\varPsi(\tau_+,\tau)
    \end{eqnarray}
In view of the equation $\mbox{\boldmath$F$}\varPsi(\tau,\tau_*)=0$
we have $(\ddot g/g)\varPsi(\tau,\tau_*)=\ddot\varPsi(\tau,\tau_*)$,
whence
    \begin{eqnarray}
    &&\delta\ddot g\;\varPsi(\tau,\tau_*)-\frac{\ddot
    g}g\,\varPsi(\tau,\tau_*)\,\delta g=
    \delta\ddot g\;\varPsi(\tau,\tau_*)
    -\delta g\,\ddot\varPsi(\tau,\tau_*)\nonumber\\
    &&\qquad\qquad\qquad\qquad\qquad\qquad\qquad\qquad
    =\frac{d}{d\tau}\,\Big(\,\delta\dot g\;\varPsi(\tau,\tau_*)
    -\delta g\,\dot\varPsi(\tau,\tau_*)\Big).
    \end{eqnarray}
Therefore
    \begin{eqnarray}
    &&\oint d\tau\,\delta\!\left(\frac{\ddot g}g\right)\,
    \left(\,\varPsi(\tau,\tau_*)\,\frac{\delta C_+}
    {\delta\tilde J(\tau)}\,\theta(\tau)
    +\varPsi(-\tau,\tau_*)\,\frac{\delta C_-}{\delta\tilde
    J(\tau)}\,\theta(-\tau)\right)\nonumber\\
    &&\qquad\qquad\quad=\frac{\varPsi_+}{\varPsi_+\dot\varPsi_+
    -\varPsi_-\dot\varPsi_-}\,
    \int_{\tau_-}^{\tau_+} d\tau\, \frac{d}{d\tau}
    \Big(\,\delta\dot g\;\varPsi(\tau,\tau_*)
    -\delta
    g\;\dot\varPsi(\tau,\tau_*)\Big)\,
    \dot\varPsi(\tau_+,\tau).                  \label{C0}
    \end{eqnarray}

Straightforward integration by parts is impossible here, because the
relevant surface terms at $\tau=\tau_\pm$ are divergent. Indeed, the
function $g(\tau)$ and its variation $\delta g(\tau)$ in view of
(\ref{conditionsong}) have expansions at $\tau=\tau_-$
    \begin{eqnarray}
    g(\tau)=\dot g_-\,\tau+O(\tau^3),\,\,\,\,
    \delta g(\tau)=\delta\dot g_-\,\tau+O(\tau^3),  \label{gatzero}
    \end{eqnarray}
and $\varPsi(\tau,\tau_*)=\varPsi_-+\dot\varPsi_-\,\tau+O(\tau^2)$,
so that $\delta\dot g\;\varPsi(\tau,\tau_*)
    -\delta
    g\;\dot\varPsi(\tau,\tau_*)=
    \delta\dot g_-\,\varPsi_-+O(\tau^2)$,
whereas
    \begin{eqnarray}
    \dot\varPsi(\tau_+,\tau)=\dot
    g_+\int_{\tau}^{\tau_+}\frac{dy}{\dot g_-^2\,y^2+...}+...
    =\frac{\dot g_+}{\dot g_-^2\,\tau}+...,\,\,\,\,\,\tau\to\tau_-= 0.
    \end{eqnarray}
Similarly
    \begin{eqnarray}
    \delta\dot g\;\varPsi(\tau,\tau_*)
    -\delta
    g\;\dot\varPsi(\tau,\tau_*)=
    \delta\dot g_+\,\varPsi_++O(\Delta^2),\,\,\,\,\,
    \dot\varPsi(\tau_+,\tau)=\frac1{\dot g_+\,\Delta},\,\,\,\,\,
    \Delta\equiv\tau-\tau_+\to 0.
    \end{eqnarray}
Consequently, the integration by parts in Eq.(\ref{C0}) is possible
only after making the subtractions of {\em constant} terms inside
the parenthesis in the right hand side of (\ref{C0}) --- $\delta\dot
g_-\,\varPsi_-$ at $\tau=\tau_-$ and $\delta\dot g_+\,\varPsi_+$ at
$\tau_+$. Thus we identically rewrite the right hand side of
(\ref{C0}) as
    \begin{eqnarray}
    &&\int_{\tau_-}^{\tau_+} d\tau\, \frac{d}{d\tau}
    \Big(\,\delta\dot g\;\varPsi(\tau,\tau_*)
    -\delta
    g\;\dot\varPsi(\tau,\tau_*)\Big)\,
    \dot\varPsi(\tau_+,\tau)\nonumber\\
    &&\qquad\qquad\qquad\quad
    =\int_{\tau_-}^{\tau_*} d\tau\, \frac{d}{d\tau}
    \Big(\,\delta\dot g\;\varPsi(\tau,\tau_*)
    -\delta
    g\;\dot\varPsi(\tau,\tau_*)-\delta\dot g_-\,\varPsi_-\Big)\,
    \dot\varPsi(\tau_+,\tau)\nonumber\\
    &&\qquad\qquad\qquad\quad
    +\int_{\tau_*}^{\tau_+} d\tau\, \frac{d}{d\tau}
    \Big(\,\delta\dot g\;\varPsi(\tau,\tau_*)
    -\delta
    g\;\dot\varPsi(\tau,\tau_*)-\delta\dot g_+\,\varPsi_+\Big)\,
    \dot\varPsi(\tau_+,\tau)                  \label{C01}
    \end{eqnarray}
and integrate by parts taking into account that
$(d/d\tau)\dot\varPsi(\tau_+,\tau)=-\dot g_+/g^2(\tau)$. This gives
    \begin{eqnarray}
    &&\int_{\tau_-}^{\tau_+} d\tau\, \frac{d}{d\tau}
    \Big(\,\delta\dot g\;\varPsi(\tau,\tau_*)
    -\delta g\;\dot\varPsi(\tau,\tau_*)\Big)\,
    \dot\varPsi(\tau_+,\tau)\nonumber\\
    &&\qquad\qquad
    =\Big(\,\delta\dot g(\tau_*)\;\varPsi(\tau_*,\tau_*)
    -\delta g(\tau_*)\;\dot\varPsi(\tau_*,\tau_*)-\delta\dot g_-\,\varPsi_-\Big)\,
    \dot\varPsi(\tau_+,\tau_*)\nonumber\\
    &&\qquad\qquad
    -\Big(\,\delta\dot g(\tau_*)\;\varPsi(\tau_*,\tau_*)
    -\delta g(\tau_*)\;\dot\varPsi(\tau_*,\tau_*)
    -\delta\dot g_+\,\varPsi_+\Big)\,
    \dot\varPsi(\tau_+,\tau_*)\nonumber\\
    &&\qquad\qquad
    +\,\dot g_+\,\Big(\,I_-+I_+\Big),                      \label{C02}
    \end{eqnarray}
where
    \begin{eqnarray}
    &&I_-=\int_{\tau_-}^{\tau_*} \frac{d\tau}{g^2(\tau)}
    \Big(\,\delta\dot g(\tau)\;\varPsi(\tau,\tau_*)
    -\delta g(\tau)\;\dot\varPsi(\tau,\tau_*)
    -\delta\dot g_-\,\varPsi_-\Big),                    \label{I-}\\
    &&I_+=\int_{\tau_*}^{\tau_+} \frac{d\tau}{g^2(\tau)}
    \Big(\,\delta\dot g(\tau)\;\varPsi(\tau,\tau_*)
    -\delta g(\tau)\;\dot\varPsi(\tau,\tau_*)
    -\delta\dot g_+\,\varPsi_+\Big).                      \label{I+}
    \end{eqnarray}
Note that the surface terms arising from integration are contributed
only by the intermediate point $\tau_*$, and they equal
    \begin{eqnarray}
    &&\Big(\,\delta\dot g(\tau_*)\;\varPsi(\tau_*,\tau_*)
    -\delta g(\tau_*)\;\dot\varPsi(\tau_*,\tau_*)
    -\delta\dot g_-\,\varPsi_-\Big)\,
    \dot\varPsi(\tau_+,\tau_*)\nonumber\\
    &&-\Big(\,\delta\dot g(\tau_*)\;\varPsi(\tau_*,\tau_*)
    -\delta g(\tau_*)\;\dot\varPsi(\tau_*,\tau_*)
    -\delta\dot g_+\,\varPsi_+\Big)\,
    \dot\varPsi(\tau_+,\tau_*)\nonumber\\
    &&\qquad\qquad\qquad\qquad\qquad\qquad\qquad\qquad
    =\left(\frac{\delta\dot g_-}{\dot g_-}
    -\frac{\delta\dot g_+}{\dot g_+}\right)\dot\varPsi_+
    =\dot\varPsi_+\,\delta\!
    \left(\ln\frac{\varPsi_+}{\varPsi_-}\right).    \label{surf0}
    \end{eqnarray}

The calculation of the integral terms with $I_\pm$ is much trickier
and is presented in Appendix B. It is based on a systematic
conversion of their integrands to the form of total derivatives in
order to generate easily calculable surface terms at
$\tau=\tau_\pm$. The result is, however, rather simple,
    \begin{eqnarray}
    &&I_-=\varPsi_-\,\delta\dot\varPsi_-
    -\delta\varPsi_-\,\dot\varPsi_-,       \label{int1}\\
    &&I_+=
    -\varPsi_+\,\delta\dot\varPsi_+
    +\delta\varPsi_+\,\dot\varPsi_+.              \label{int2}
    \end{eqnarray}

Collecting (\ref{surf0}), (\ref{int1}), (\ref{int2}) and (\ref{C02})
we get the total contribution of C-terms (\ref{C}).

\section{The contribution of integral terms with $I_\pm$}
The strategy of calculating the integrals (\ref{I-})-(\ref{I+})
consists in a systematic conversion of their integrands to the form
of total derivatives which yield easily calculable surface terms at
$\tau=\tau_\pm$. For this purpose we use the corollaries of equation
(\ref{Psi}) for $\varPsi=\varPsi(\tau,\tau')$,
    \begin{eqnarray}
    &&\dot\varPsi=\frac{\dot g}g\,\varPsi+\frac1g\,, \label{dotPsi}\\
    &&\frac1{g^2}=\frac{d}{d\tau}
    \left(\frac{\varPsi}g\right)\,.
    \end{eqnarray}
In view of the last equation the last term in the integrand of $I_-$
is the total derivative
    \begin{eqnarray}
    -\frac{d}{d\tau}\left(\frac{\varPsi}g\,\varPsi_-\delta\dot g_-\right)
    =\frac{d}{d\tau}\left(\frac{\varPsi}g\,\frac{\delta\dot g_-}{\dot g_-}\right),
    \end{eqnarray}
whereas the first two terms can be transformed by using
(\ref{dotPsi}) as
    \begin{eqnarray}
    \frac1{g^2}
    \Big(\,\delta\dot g\;\varPsi
    -\delta g\;\dot\varPsi
    \Big)=\frac{d}{d\tau}\left(\frac{\delta g\,\varPsi}{g^2}\right)
    +\frac{2\delta g\,\dot g}{g^3}\,\varPsi
    -\frac{2\delta g}{g^2}\,\dot\varPsi
    =\frac{d}{d\tau}\left(\frac{\delta g\,\varPsi}{g^2}\right)
    -\frac{2\delta g}{g^3},
    \end{eqnarray}
whence
    \begin{eqnarray}
    &&I_-
    =\int_{\tau_-}^{\tau_*}
    d\tau\left\{\frac{d}{d\tau}\left[\frac\varPsi{g}
    \left(\frac{\delta g}g+\frac{\delta\dot g_-}{\dot
    g_-}\right)\right]
    -\frac{2\delta g}{g^3}\right\}.              \label{integral}
    \end{eqnarray}

The integrand is not yet a total derivative. Moreover, the total
derivative term cannot be reduced to the surface term, because the
latter diverges at $\tau=\tau_-$ as $-2\delta\dot g_-/\dot
g_-^3\,\tau$, $\tau\to 0$ (remember that $\varPsi(\tau_-)=-1/\dot
g_-$, $g(\tau)\sim\dot g_-\,\tau$ and $\delta g(\tau)\sim\delta\dot
g_-\,\tau$). This divergence is in fact canceled in the integrand by
the last term $-2\delta g/g^3\sim -2\delta\dot g_-/\dot
g_-^3\,\tau^2$. Thus, these two terms cannot be treated separately,
because of this subtraction mechanism. Instead of splitting them we
will use another quantity in which $\delta g/g^3$ also enters the
integrand. This quantity is a variation of the expression
    \begin{eqnarray}
    &&\varPsi\,\dot\varPsi\Big|_{\,\tau=\tau_-}^{\,\tau=\tau_*}=
    \int_{\tau_-}^{\tau_*}
    d\tau\frac{d}{d\tau}\big(\varPsi\,\dot\varPsi\big)=
    \int_{\tau_-}^{\tau_*}
    d\tau\,\left(\dot\varPsi^2+\varPsi\,\ddot\varPsi\right)=
    \int_{\tau_-}^{\tau_*}
    d\tau\,\Big(\dot\varPsi^2+\frac{\ddot g}g\,\varPsi^2\Big),
    \end{eqnarray}
in which we used the equation for $\varPsi$, $\ddot\varPsi=(\ddot
g/g)\varPsi$. Using (\ref{dotPsi}) to express $\dot\varPsi$ in terms
of $\varPsi$ one easily shows that
    \begin{eqnarray}
    \varPsi\,\dot\varPsi\Big|_{\,\tau=\tau_-}^{\,\tau=\tau_*}=
    \int_{\tau_-}^{\tau_*}
    d\tau\,\left\{\frac{d}{d\tau}
    \left(\frac{\dot g}g\,\varPsi^2\right)+\frac1{g^2}\right\}.
    \end{eqnarray}
The variation of this expression has the same structure as
(\ref{integral})
    \begin{eqnarray}
    \delta\left(\varPsi\,\dot\varPsi\Big|_{\,\tau_-}^{\,\tau_*}\right)=
    \int_{\tau_-}^{\tau_*}
    d\tau\,\left\{\frac{d}{d\tau}
    \delta\!\left(\frac{\dot g}g\,\varPsi^2\!\right)
    -\frac{2\delta g}{g^3}\right\}.
    \end{eqnarray}
Therefore, the subtraction of this expression from (\ref{integral})
allows one to exclude the integral of $-2\delta g/g^3$ and to obtain
    \begin{eqnarray}
    &&I_- -\delta\left(\varPsi\,\dot\varPsi\Big|_{\,\tau_-}^{\,\tau_*}\right)
    =\int_{\tau_-}^{\tau_*}
    d\tau\,\frac{d}{d\tau}\left[\frac\varPsi{g}
    \left(\frac{\delta g}g+\frac{\delta\dot g_-}{\dot
    g_-}\right)
    -\delta\!\left(\frac{\dot
    g}g\,\varPsi^2\!\right)\right]\nonumber\\
    &&\qquad\qquad\qquad\quad\quad=\left[\,
    \delta\Big(\,\frac{\dot g}g\,
    \varPsi^2\Big)-\frac\varPsi{g}
    \Big(\frac{\delta g}g+\frac{\delta\dot g_-}{\dot
    g_-}\Big)\,\right]_{\,\tau=\tau_-},              \label{integral1}
    \end{eqnarray}
where the surface term at $\tau_*$ vanishes in view of
$\varPsi_*\equiv\varPsi(\tau_*,\tau_*)=0$ and the surface term at
$\tau_-$ is now finite, because of the cancelation of divergences
between the two terms in the last line of the above equation. To
find the remnant of this cancelation we write
    \begin{eqnarray}
    &&\left[\,
    \delta\Big(\,\frac{\dot g}g\,
    \varPsi^2\Big)-\frac\varPsi{g}
    \Big(\frac{\delta g}g+\frac{\delta\dot g_-}{\dot
    g_-}\Big)\,\right]_{\,\tau=\tau_-}\nonumber\\
    &&\qquad\qquad
    =\frac1g\,\left[\,\delta\dot g\,
    \varPsi^2-\frac{\delta g\,\dot g}{g}\,
    \varPsi^2+2\dot g\,
    \varPsi\,\delta\varPsi
    -\varPsi
    \Big(\frac{\delta g}g+\frac{\delta\dot g_-}{\dot
    g_-}\Big)\,\right]_{\,\tau=\tau_-}\nonumber\\
    &&\qquad\qquad
    =\frac1{\dot g_-}\frac{d}{d\tau}
    \left[\,\delta\dot g\,
    \varPsi^2-\frac{\delta g\,\dot g}{g}\,
    \varPsi^2+2\dot g\,
    \varPsi\,\delta\varPsi
    -\varPsi
    \Big(\frac{\delta g}g+\frac{\delta\dot g_-}{\dot
    g_-}\Big)\,\right]_{\,\tau=\tau_-}
    =2\varPsi_-\,
    \delta\dot\varPsi_-,                    \label{surfterm}
    \end{eqnarray}
where we took into account that in view of (\ref{gatzero})
$\delta\ddot g_-=\ddot g_-=0$, $(d/d\tau)(\delta g/g)|_0=0$ and
$\delta\varPsi_-=-\delta(1/\dot g_-)$. Therefore, because
$\varPsi(\tau_*,\tau_*)=\delta\varPsi(\tau_*,\tau_*)=0$, we finally
have the expression (\ref{int1}) for $I_-$ used above.

A similar calculation for the second integral term gives the
equations analogous to (\ref{integral1}) and (\ref{surfterm})
    \begin{eqnarray}
    &&I_+-
    \delta\left(\varPsi\,
    \dot\varPsi\Big|_{\,\tau_*}^{\,\tau_+}\right)
    =-\left[\,\delta\Big(\frac{\dot g}g\,
    \varPsi^2\Big)-\frac\varPsi{g}
    \Big(\frac{\delta g}g+\frac{\delta\dot g_+}{\dot
    g_+}\Big)\,\right]_{\,\tau_+}=-2\varPsi_+\,
    \delta\dot\varPsi_+,              \label{integral2}
    \end{eqnarray}
except the opposite sign caused by the upper integration limit
(rather than by the lower one). This leads to (\ref{int2}).

\section{The contribution of the D-terms}
By using (\ref{halfofD+D}) one has
    \begin{eqnarray}
    &&Q\,\left.\frac{D_++D_-}2\,
    \right|_{\,\textstyle\tilde J=\delta(\ddot g/g)\,g(\tau)}
    =\oint dy\,g(y)\,\int_{\tau_-}^y d\tau\,\varPsi(y,\tau)\,
    \frac{d}{d\tau}(\,g\,\delta\dot g-\dot g\,\delta g)\nonumber\\
    &&\qquad\qquad\qquad\qquad\qquad\qquad\quad
    =\oint dy\,g^2(y)\,\int_{\tau_-}^y
    d\tau\,\frac{d}{d\tau}\left(\frac{\delta g}g\right),
    \end{eqnarray}
where when integrating by parts no surface terms arise, because
$\varPsi(y,y)=0$ and $(\,g\,\delta\dot g-\dot g\,\delta
g)=O(\tau^2)$ at $\tau\to 0$, and we used the fact that
$(d/d\tau)\varPsi(y,\tau)=-g(y)/g^2(\tau)$. Thus
    \begin{eqnarray}
    &&\left.\frac{D_++D_-}2\,
    \right|_{\,\textstyle\tilde J=\delta(\ddot g/g)\,g(\tau)}
    =\frac12\,\delta\ln Q-
    \frac{\delta g}g(\tau_-).               \label{D01}
    \end{eqnarray}
Here we perform the variation of $g(\tau)$ in the class of functions
satisfying $\delta g(\tau)=\delta\dot g_-\,\tau+O(\tau^2)$, so that
with $g(\tau)=\dot g_-\,\tau+O(\tau^2)$ we have $(\delta
g/g)(\tau_-)=\delta\dot g_-/\dot g_-=-\delta\varPsi_-/\varPsi_-$,
which finally leads to (\ref{D1}).

The second term in Eq. (\ref{Dgeneral}) in view of (\ref{D-D}) and
(\ref{dotPhipm}) reads
    \begin{eqnarray}
    &&\left.\frac{D_+-D_-}2\,
    \right|_{\,\textstyle\tilde J
    =\delta(\ddot g/g)\,g\,
    \varepsilon(\tau)}=
    \frac12\frac{\varPsi_+\varPsi_-\dot\varPsi_-}
    {\varPsi_+\dot\varPsi_+
    -\varPsi_-\dot\varPsi_-}\,\left(
    \int_{\tau_-}^{\tau_+} dy\,\dot\varPsi(\tau_+,y)\,
    g^2\delta\!\left(\frac{\ddot g}g\right)(y)\right.\nonumber\\
    &&\qquad\qquad\qquad\qquad\qquad\qquad\qquad\qquad\qquad
    +\left.
    \int_{\tau_-}^{-\tau_+} dy\,\dot\varPsi(-\tau_+,y)\,
    g^2\delta\!\left(\frac{\ddot g}g\right)(y)\right).         \label{D-Dcontribution}
    \end{eqnarray}
In view of the symmetries of the functions in the integrand,
$\dot\varPsi(-\tau_+,y)=-\dot\varPsi(\tau_+,-y)$, $g(y)=-g(-y)$ and
$\delta(\ddot g/g)(-y)=\delta(\ddot g/g)(y)$ the two integrals above
coincide and
    \begin{eqnarray}
    &&\left.\frac{D_+-D_-}2\,
    \right|_{\,\textstyle\tilde J
    =\delta(\ddot g/g)\,g\,
    \varepsilon(\tau)}
    =\frac{\varPsi_+\varPsi_-\dot\varPsi_-}
    {\varPsi_+\dot\varPsi_+
    -\varPsi_-\dot\varPsi_-}\,
    \int_{\tau_-}^{\tau_+} dy\,\dot\varPsi(\tau_+,y)\,
    \frac{d}{dy}(g\,\delta\dot g-\dot g\,\delta g).
    \end{eqnarray}
Here integration by parts does not give surface terms because
$g\,\delta\dot g-\dot g\,\delta g=O(\Delta^2)$,
$\dot\varPsi(\tau_+,y)=O(1/\Delta)$ at $\Delta\equiv y-\tau_\pm\to
0$. Therefore
    \begin{eqnarray}
    &&\left.\frac{D_+-D_-}2\,
    \right|_{\,\textstyle\tilde J
    =\delta(\ddot g/g)\,g\,
    \varepsilon(\tau)}
    =-\frac{\varPsi_+\varPsi_-\dot\varPsi_-}
    {\varPsi_+\dot\varPsi_+
    -\varPsi_-\dot\varPsi_-}\,
    \int_{\tau_-}^{\tau_+} dy\,\frac{d}{dy}\dot\varPsi(\tau_+,y)\,
    (g\,\delta\dot g-\dot g\,\delta g)\nonumber\\
    &&\qquad\qquad\qquad\qquad\qquad\qquad\quad
    =\frac{\varPsi_-\dot\varPsi_-}
    {\varPsi_+\dot\varPsi_+
    -\varPsi_-\dot\varPsi_-}\,
    \left(\frac{\delta \varPsi_+}{\varPsi_+}-
    \frac{\delta \varPsi_-}{\varPsi_-}\right),  \label{D02}
    \end{eqnarray}
where we took into account that $(d/dy)\dot\varPsi(\tau_+,y)=-\dot
g_+/g^2(y)$ and $\delta g/g(\tau_+)=\delta\dot g_+/\dot g_+$ and
$\delta g/g(0)=\delta\dot g_-/\dot g_-$. This gives (\ref{D2}).

\end{document}